\documentclass[mypaper,7pt,twoside]{CoAst}
\usepackage{epsf,graphicx,fancyhdr}
\renewcommand{\chaptermark}[1]{\markboth{#1}{}}

\newcommand{\kopf}{\small\itshape Comm. in Asteroseismology \\ Contribution to the Proceedings of the Wroclaw HELAS Workshop, 2008}

\newcommand{\Authors}[1]{\begin{center}\normalsize\bf\sf #1 \end{center}}

\renewcommand{\author}[1]{\begin{center}\normalsize\bf\sf #1 \end{center}}
\newcommand{\Address}[1]{\begin{center}\small\sf #1 \end{center}}

\newcommand{\Session}[1]{{\vspace{3mm}\small \noindent  \hspace*{3mm} Session: } #1 \normalsize}

\newcommand{\Objects}[1]{{\vspace{0mm}\small \noindent  \hspace*{3mm} Individual Objects: } \small #1 \normalsize}

	\newcommand{\threeD}{\small STARS - effects of magnetic field on stellar pulsation \newline}

\renewenvironment{abstract}{\section*{Abstract}\normalsize\sf}{}
\newcommand{\References}[1]{\begin{flushleft}{\large References\\}\vspace*{2mm}\small #1 \end{flushleft}}

\newcommand{\chapterCoAst}[2]{\chapter[\sf\normalsize #1\\ \footnotesize \hspace*{5mm}by #2 \sf\normalsize][]{#1\\}\rhead[\fancyplain{}{\sf\footnotesize \center{#1}}]{\fancyplain{}{\sffamily\thepage}}\lhead[\fancyplain{\kopf}{\sffamily\thepage}]{\fancyplain{\kopf}{\sf\footnotesize \center{#2}}}}

\renewcommand{\chaptername}{}

\def\rfr{\smallskip\par\noindent
        \hangindent=7truemm
        \hangafter=1}

\begin{document}
\sf
\chapterCoAst{Interferometric and seismic constraints \\ on the roAp star 
$\alpha$\,Cir}
{I.\,M.\,Brand\~ao, H.\,Bruntt, M.\,Cunha, D.\,W.\,Kurtz} 
\Authors{I.\,M.\,Brand\~ao$^{1,2}$, H.\,Bruntt$^{3}$, M.\,Cunha$^1$, 
D.\,W.\,Kurtz$^4$} 
\Address{
$^1$ Centro de Astrof\'isica, Universidade do Porto, \\
     Rua das Estrelas, 4150-762 Porto, Portugal\\
$^2$ Departamento de Matem\'atica Aplicada, \\
     Faculdade de Ci\^encias da Universidade do Porto, Portugal\\
$^3$ Institute of Astronomy, School of Physics A28, University of Sydney, Australia\\
$^4$ Centre for Astrophysics, University of Central Lancashire, Preston, U.K.
}
\noindent
\begin{abstract}
We present new constraints on the rapidly oscillating Ap star $\alpha$\,Cir, 
derived from a combination of interferometric and photometric data obtained with 
the Sydney University Stellar Interferometer (SUSI) and the \textit{WIRE} 
satellite. The highlights of our study are:

\begin{enumerate}

\item The first determination of the angular diameter of an roAp star.

\item A nearly model-independent determination of the effective temperature of $\alpha$\,Cir, 
which is found to be lower than previously estimated values.

\item Detection of two new oscillation frequencies allowing 
a determination of the large separation of $\alpha$\,Cir.

\end{enumerate}

Based on this new information, we have computed non-magnetic and magnetic models 
for $\alpha$\,Cir. We show that the value of the observed large separation 
found from the new data agrees well with that derived from theoretical models. 
Moreover, we also show how the magnetic field may explain some of the anomalies 
seen in the oscillation spectrum and how these in turn
provide constraints on the magnitude and topology of the magnetic field.
\end{abstract}

\Session{\threeD}
\Objects{$\alpha$\,Cir -- HD~128898} 
\section*{Introduction}

$\alpha$\,Circini [HR\,5463, HD\,128898, HIP\,71908, $V = 3.2$] is the brightest 
known rapidly oscillating peculiar A-type (roAp) star. The roAp stars are main-sequence 
chemically peculiar (CP) pulsators with effective temperatures ranging from
6\,500 to 8\,500\,K. Since CP stars show abnormal flux distributions in their 
spectra, their effective temperatures are very difficult to determine. 
Temperatures can be estimated from photometric indices or spectral analysis, 
but due to the peculiar nature of these stars, 
values are likely to be affected by systematic effects. 

The roAp stars also present the highest oscillation frequencies observed in 
the main sequence part of the instability strip, with typical values 
ranging from 1 to 3\,mHz. The high frequencies of the oscillations observed in roAp 
stars indicate that these are high radial order, low degree acoustic modes. Since 
the oscillations are of high radial order we can, in principle, use the asymptotic 
theory to study the oscillation spectrum. However, these oscillations are affected 
by an intense magnetic field that will 
perturb the frequencies from the asymptotic trend.

$\alpha$\,Cir is one of the best studied roAp stars and, as such, both seismic and 
non-seismic data for this star are available in the literature. However, to date, 
the large frequency separation (defined as the difference between the frequencies 
of modes of the same degree and consecutive radial orders) of $\alpha$\,Cir cannot 
be reconciled with that expected from an effective temperature around 8\,000\,K, 
suggested by most determinations found in the literature, and the luminosity 
derived from the \textit{Hipparcos} parallax 
(Matthews et al.\ 1999).

\section*{Detection of the large separation}

\begin{figure}[tbhp]
\centering
\includegraphics[scale=0.7]{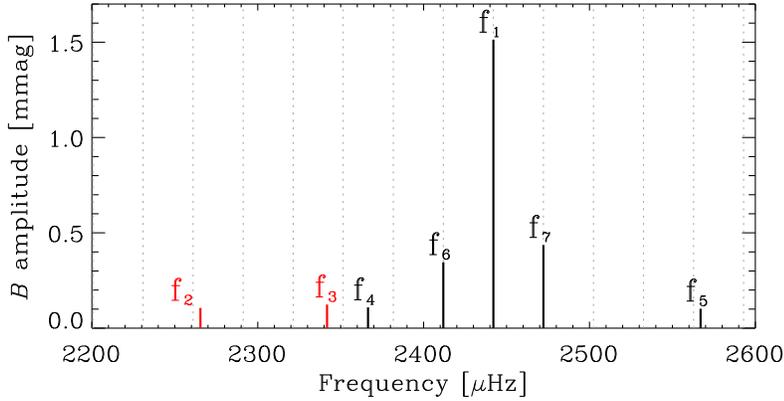}
\caption{\small  Frequencies detected in $\alpha$\,Cir from the \textit{WIRE} data, $f_4$, 
$f_6$, $f_1$, $f_7$, $f_5$. We have included $f_2$ and $f_3$ from Kurtz et al.\ 
(1994). The vertical dashed lines mark half the large separation (the mean of  
$f_1 - f_6$ and $f_7 - f_1$).  \normalsize}
\label{fig:obsoscfreq}
\end{figure}

$\alpha$\,Cir was observed for 84\,d during four runs 
with the \textit{WIRE} satellite in the period from 2000--2006 
(Bruntt et al.,\ private communication).
During the last two runs we collected simultaneous 
ground-based Johnson $B$ observations on 16 nights with the 0.5-m and 
0.75-m telescopes at the South African Astronomical Observatory (SAAO) and 2\,hr 
of high-cadence, high-resolution spectra from the Ultraviolet and Visual Echelle 
Spectrograph (UVES) on the Very Large Telescope (VLT). 
The oscillation frequencies detected
in the \textit{WIRE} data are shown in Figure~\ref{fig:obsoscfreq}.
The $f_6$ and $f_7$ frequencies have not been observed before, 
and are present in both the \textit{WIRE} and SAAO data sets.
The $f_6+f_1+f_7$ frequencies have the highest amplitudes and form
a triplet with a nearly equidistant frequency spacing of $30.173 \pm0.004$\,$\mu$Hz. 
We interpret this spacing as either the large frequency separation or half of that.

\section*{Asteroseismology}

\subsection*{Non-magnetic model}

\begin{figure}[!ht]
\centering
\includegraphics[scale=0.35]{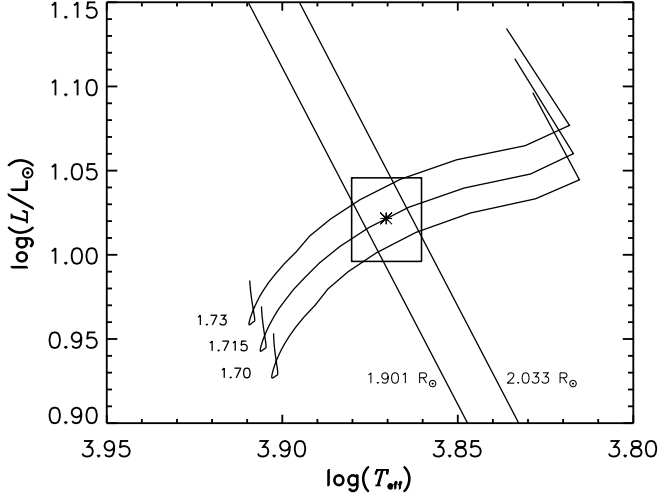}
\caption{\small The position of $\alpha$\,Cir in the HR diagram, with three 
evolutionary tracks for masses of 1.70, 1.715 and 1.73 M$_\odot$. The 
constraints on the fundamental parameters are indicated by the 1-$\sigma$ error 
box ($T_{\rm  eff}$, $L$/L$_\odot$) and the diagonal lines (radius).\normalsize}
\label{fig:acirHR}
\end{figure}

Bruntt et al.\ (2008) determined the effective temperature of $\alpha$\,Cir by 
combining the measured angular diameter of the star obtained with the Sydney 
University Stellar Interferometer (SUSI) and its bolometric flux, computed from 
calibrated spectra. They found a nearly model-independent 
value for the effective temperature of $7420\pm170$\,K, 
which is lower than all previous determinations found in the literature. 
The new values for the effective temperature and luminosity, derived 
from the \textit{Hipparcos} parallax and the interferometric radius, were used to 
place $\alpha$\,Cir in the Hertzsprung-Russell (HR) diagram 
as shown in Figure~\ref{fig:acirHR}.

Three CESAM (Morel 1997) evolutionary tracks that go through the 1-$\sigma$ error box 
are shown in Figure~\ref{fig:acirHR}. We chose the model that best fitted 
the position of the star in the HR~diagram and its parameters 
are given in Table~\ref{tab:model}. 
We calculated the theoretical oscillation frequencies for that 
model with the linear adiabatic oscillation code Aarhus Adiabatic Pulsation 
Package (ADIPLS; Christensen-Dalsgaard 2008). From these theoretical frequencies 
we calculated the large frequency separation and obtained a value of $\Delta \nu = 
60.4\,\mu$Hz. Comparing this value with the observed frequency spacing we 
conclude that this model reproduces well the observed separation between the three 
principal modes. Moreover, we conclude that the 
observed large frequency separation of $\alpha$\,Cir is $60.346 \pm 
0.008$\,$\mu$Hz and, thus, that the frequencies $f_6$, $f_1$ and $f_7$ must 
correspond to modes of alternating even-odd spherical degrees. This new value is 
significantly larger -- and much more secure -- than the $50$\,$\mu$Hz 
suggested by Kurtz et al.\ (1994).

\begin{table}[h]
\caption{\small Global parameters of the CESAM model used for $\alpha$\,Cir. The 
following input parameters were used: $X_0= 0.70$, $Y_0=0.28$, $\alpha=1.6$ and no 
overshooting. $X_0$ and $Y_0$ are the initial H and He abundances 
and $\alpha$ is the mixing length parameter.} 
\label{tab:model} \centering
\begin{tabular}{l|ccccc}     \hline \hline
 & $M$/M$_\odot$  &  log($L$/L$_\odot$)   &  $\log T_{\rm eff}$ [K] & 
$R$/R$_\odot$ & Age (Myr)\\ \hline
CESAM model  & 1.715  & 1.022  & 3.87  &   2.0 & 900 \\
\hline 
\end{tabular} 
\end{table}

\subsection*{Magnetic models}

Inspecting Figure~\ref{fig:obsoscfreq}, where the spacing between the dashed 
vertical lines correspond to half of the large separation, we note that only the 
three principal modes seem to follow the trend expected in the asymptotic regime. 
In particular, $f_4$, which was also observed by Kurtz et al.\ (1994), is separated 
from $f_6$ by $\simeq3/4$ of the large separation $(f_6-f_4=45.41~\mu$Hz). Consequently, 
the oscillation frequencies computed with the model in Table 
\ref{tab:model} (hereafter called the {\it non-magnetic model}) do not reproduce 
well the separation between the principal mode and the frequencies $f_2$, $f_3$, 
$f_4$ and $f_5$. Moreover, the frequencies of the three principal 
modes have nearly equal separations:
$(f_1-f_6)=30.1746\pm0.0009\,\mu$Hz and 
$(f_7-f_1)=30.1707\pm0.0005\,\mu$Hz. 
In fact, the difference 
between these two ``half separations'', which we will denominate by $\delta\nu_{\rm obs}$, is 
only $0.004\pm0.001\,\mu$Hz. After computing theoretical $\delta\nu_{nl}$ values
for all combinations of mode degrees with $l \le 3$ for the non-magnetic model, 
we found that the minimum 
absolute value taken by this quantity is $\delta\nu_{nl}=2.5\,\mu$Hz. This value is 
obtained for combinations of modes of degree $l=0$ and $2$, around the frequency 
$2450\,\mu$Hz.  


Since $\alpha$\,Cir is an roAp star, it has a strong magnetic field. 
We have therefore speculated if the effect of the magnetic field 
on the oscillations may explain the small value of $\delta\nu_{\rm obs}$.
To investigate this possibility, we used a code 
(Cunha 2006) to compute the magnetic perturbations to the frequencies 
obtained for our non-magnetic model. 
As input parameters we considered modes of degrees 
$l=0,1,2$ and $3$, a magnetic field at the pole, $B_p$, within a range of values 
appropriate for $\alpha$\,Cir, 
(see Bruntt et al.\ 2008: Sec. 6.1, for a review)
and a dipolar or quadrupolar magnetic field topology. 
The three magnetic models that best reproduce the features of the 
oscillation spectra of $\alpha$\,Cir are shown in 
Table~\ref{tab:bestmagmodel} and the values of $l$ that correspond to each 
frequency for these models are given in Table~\ref{tab:bestmagmodell}.

\begin{table}
\caption{\small The values of $\delta \nu_{nl}$ and $(f_6-f_4)$ 
for the observations and
for the best-fitting non-magnetic and magnetic models.
The strength and topology of the magnetic field are given for the three magnetic models.\normalsize} 
\label{tab:bestmagmodel} \centering
\begin{tabular}{l|cllc}     \hline \hline
  & $B_p$  &  \multicolumn{1}{c}{Topology}   & \multicolumn{1}{c}{$\delta \nu_{nl}$} & $(f_6-f_4)$ \\ 
                   &[kG]    &              & [$\mu$Hz]         & [$\mu$Hz]   \\ \hline
Observed values    & $-$    & $-$          & $+0.004$          & $45.41$     \\ \hline
Non-magnetic model & $-$    & $-$          & $+2.5  $          & $30.2$     \\
Magn.\ model 1     & $1.4$  & Quadrupolar  & $-0.66 $          & $44.76$     \\
Magn.\ model 2     & $1.4$  & Quadrupolar  & $+0.53 $          & $40.57$     \\
Magn.\ model 3     & $1.4$  & Dipolar      & $-0.81 $          & $50.81$     \\
\hline 
\end{tabular} 
\end{table}

\begin{table}
\caption{\small For each of the three best magnetic models and for the best non-magnetic model
we list the values of $l$ for the four frequencies $f_4$, $f_6$, $f_1$, and $f_7$. \normalsize} 
\label{tab:bestmagmodell} \centering
\begin{tabular}{l|cccc}     \hline \hline
 & $l_{f_4}$ & $l_{f_6}$ & $l_{f_1}$ & $l_{f_7}$ \\ \hline
Non-magnetic model & 1  & 0 & 1 & 0\\
Magn.\ model 1     & 3  & 2 & 3 & 2\\
Magn.\ model 2     & 1  & 3 & 2 & 3\\
Magn.\ model 3     & 0  & 2 & 3 & 2\\
\hline 
\end{tabular} 
\end{table}

\section*{Conclusions and discussion}

We have summarized the main results of an intensive study of the roAp star 
prototype $\alpha$\,Cir, part of which has been published in 
Bruntt et al.\ 2008. 
Our team has made the first interferometrically-based determination 
of the effective temperature of an roAp star.
The new value of $T_{\rm eff}=7420\pm170$\,K 
is lower than all values found in the literature. 
Additionally, new seismic data for $\alpha$\,Cir were acquired 
with the \textit{WIRE} satellite and with the 0.5-m and 0.75-m telescopes at SAAO.
Two new frequencies were found in both the \textit{WIRE} and SAAO data
and they form a triplet with the known dominant frequency. 
The triplet is nearly equally spaced with a separation of $30.173 \pm0.004$\,$\mu$Hz, 
which we interpret to be half the large separation. 
Using the new global parameters of the star, we computed
a non-magnetic model for $\alpha$\,Cir. The large separation of this model is
in good agreement with the observed large separation, 
but the model fails to explain the nearly equidistant spacing 
as well as the secondary frequencies.

In an attempt to understand these discrepancies we computed magnetic perturbations 
to the frequencies of the non-magnetic model. We found that the magnetic model 
that best reproduces the oscillation spectrum has a quadrupolar topology and a 
magnitude of $1.4\,$kG. From this model, we identify the largest amplitude mode, 
$f_1$, as being an $l=3$ mode. We note that due to the magnetic effect, the
eigenfunctions in roAp stars are distorted. Thus, it is possible that modes of 
degree higher than l = 2 may generate lower-degree components near the surface 
that, in turn, may be observed (e.g. Cunha 2005). Also, we find that the magnitude is rather sensitive to 
the position where one of the 
boundary conditions of the magnetic code is applied. To overcome this problem, and 
test the robustness of our results, we are currently implementing a different 
atmospheric model in our code. 
Thus, the results presented here for the magnetic models are still preliminary.

\References{
\rfr Bruntt H., North J. R., Cunha M., et al.\ 2008, MNRAS, 386, 2039
\rfr Christensen-Dalsgaard, J. 2008, Ap\&SS, 2008, CoRoT/ESTA Vol.
\rfr Cunha M. S. 2005, JApA, 26, 213
\rfr Cunha M. S. 2006, MNRAS, 365, 153
\rfr Kurtz D. W., Sullivan D. J., Martinez P., \& Tripe P. 1994, MNRAS , 270, 674
\rfr Matthews J. M., Kurtz D. W., \& Martinez P. 1999, ApJ, 511, 422 
\rfr Morel P. 1997, A\&AS, 124, 597
}

\end{document}